\documentclass[final,3p,times,twocolumn]{elsarticle}
\usepackage{graphicx}
\usepackage{amssymb}
\usepackage{amsthm}

\def\beq{\begin{eqnarray}}    %%  begequation/eqnarray
\def\eeq{\end{eqnarray}}      %%  endequation/eqnarray

\journal{Physics Letters B}

\begin{document}

\begin{frontmatter}

%% Title, authors and addresses

%% use the tnoteref command within \title for footnotes;
%% use the tnotetext command for the associated footnote;
%% use the fnref command within \author or \address for footnotes;
%% use the fntext command for the associated footnote;
%% use the corref command within \author for corresponding author footnotes;
%% use the cortext command for the associated footnote;
%% use the ead command for the email address,
%% and the form \ead[url] for the home page:
%%
%% \title{Title\tnoteref{label1}}
%% \tnotetext[label1]{}
%% \author{Name\corref{cor1}\fnref{label2}}
%% \ead{email address}
%% \ead[url]{home page}
%% \fntext[label2]{}
%% \cortext[cor1]{}
%% \address{Address\fnref{label3}}
%% \fntext[label3]{}

\title{On tunneling across horizons}

%% use optional labels to link authors explicitly to addresses:
%% \author[label1,label2]{<author name>}
%% \address[label1]{<address>}
%% \address[label2]{<address>}

\author{Luciano Vanzo}

\address{Dipartimento di Fisica,  Universit\`a di Trento
and Istituto Nazionale di Fisica Nucleare,\\
 Gruppo Collegato di Trento, Via Sommarive 14,
I-38023 Povo (TN),\\ Italia}

\begin{abstract}
The tunneling method for stationary black holes in the Hamilton-Jacobi variant is reconsidered in the light of various critiques that have been moved against. It is shown that once the tunneling trajectories have been correctly identified the method is free from internal inconsistencies, it is manifestly covariant, it allows for the extension to spinning particles and it can even be used without solving the Hamilton-Jacobi equation. These conclusions borrow support on a simple analytic continuation of the classical action of a pointlike particle, made possible by the unique assumption that it should be analytic in complexified Schwarzschild or Kerr-Newman spacetimes. A more general version of the Parikh-Wilczek method will also be proposed along these lines.
\end{abstract}

\begin{keyword}
Black holes \sep Hawking radiation \sep quantum tunneling

%% MSC codes here, in the form: \MSC code \sep code
%% or \MSC[2008] code \sep code (2000 is the default)

\end{keyword}

\end{frontmatter}

%%
%% Start line numbering here if you want
%%
% \linenumbers

%% main text
\section{Introduction}
\label{1:Intr}

When the tunneling method was first proposed \cite{Parikh:1999mf,padma:1999,padma:2001,padma:2002,padma:2004,Visser:2001kq,Angheben:2005rm,Arzano:2005rs,Medved:2005yf} a certain discomfort appeared soon, having mainly to do with two aspects of the method. One was that even employing a coordinate system covering regularly the horizon, the action (or the radial momentum) exhibited a pole, thus demanding a proper treatment; on the other hand, a direct integration across the singularity in Schwarzschild  coordinates (or Boyer-Lindquist for Kerr) produced twice the correct Hawking's temperature of the BH, as noted in \cite{Angheben:2005rm}. The second had to do with the possibility that a complex contribution from the temporal part of the classical action $I$, namely the term $\int\partial_tIdt$,  could either cancel or doubling the relevant emission term, if not properly handled. Several proposal where soon advanced \cite{Angheben:2005rm,Mitra:2006qa,Stotyn:2008qu,Akhmedova:2008dz,deGill:2010nb,Zhang:2009yma,Wu:2007ty}.  Although each one has its own merits, we shall see that no one is particularly compelling.  Most of the derivations and the ensuing problems have to do with coordinates choices, or lack of manifest general covariance\footnote{Even the top cited paper by Parikh \& Wilczek \cite{Parikh:1999mf} used Painlev\'e-Gullstrand coordinates, thus avoiding problems with covariance.} (see \cite{Wu:2007ty} for a particular approach to this problem for weakly isolated horizons). To be completely clear it must be said that when Parikh \& Wilczek \cite{Parikh:1999mf} (see also \cite{Parikh:2004ih}) introduced their method, successively dubbed "the null geodesics method",  the main motivation was to reveal the back reaction corrections to first order in $\epsilon/M$, where $M$ is the black hole mass and $\epsilon$ the energy scale of the process.  No back reaction correction will be considered here, as it is not really needed to derive the Hawking effect. Our emphasis will be on covariance with the aim of resolving certain conflicts between different views, hence the preferred method should be invariant ab initio.  To this aim we will work with the Hamilton-Jacobi (abbr. HJ) version of the method.\\
So after a brief review of this method we shall propose such a covariant derivation of the tunneling method. A "covariantized" Parikh-Wilczek method will also be proposed along these lines. 

\section{The tunneling path and the method} 
According to a standard picture of the black hole radiation, a pair is created somewhere near the horizon, one member escaping to infinity with positive energy, the other falling down the black hole and carrying negative energy. And this process continuously occurring here and there adds up to form the thermal streaming from the black hole. According to Hartle and Hawking \cite{Hartle:1976tp}, this can be described equivalently as the escape of a particle along the forbidden path, the right one in the figure. We shall call this the tunneling path.  Once out of the horizon the particle cannot do anything but moving forward in time, eventually ending its journey reflected back into the black hole.

\begin{figure}[h]
\includegraphics[scale=0.5]{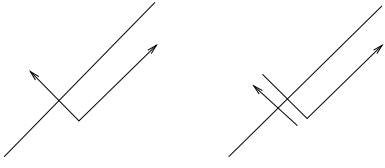}
\caption{To the left, a pair come into being near the horizon; to the right, part of the equivalent tunneling path and a segment of an ingoing path.}
\end{figure}
To see what motivated the perplexities, let us review the HJ method starting  with the Schwarzschild metric
\beq\label{sch}
ds^2=-\left(1-\frac{2m}{r}\right)dt^2+\left(1-\frac{2m}{r}\right)^{-1}dr^2+r^2d\omega^2
\eeq 
In the tunneling method one is interested in the imaginary part of the integral of the action along the tunneling path, say
\begin{equation}
\Im\int\!\!\!\!\!\!\!\searrow dI 
\end{equation}
the arrow denoting the horizon crossing path  and $dI$  the differential of the action of a spinless massless particle. Ignoring for the time being that the portion of the path crossing the horizon is not covered by a single coordinate patch, the HJ equation gives
\beq\label{action}
I=-Et+\int^r \frac{E rdr}{r-2m} + J\phi
\eeq
It is argued that the internal segment can only be reached through a journey into complex $r$-space, and the right procedure to do so, as will be seen, is Feynman $i\epsilon$-prescription\footnote{This was first introduced in \cite{Damour:1976jd} and justified without using Kruskal coordinates in \cite{padma:1999}. Extension to Lemaitre and Painlev\'e coordinates were presented in \cite{padma:2001} and taken to indicate the covariance we alluded for. }: the substitution $r-2m\to r-2m-i\epsilon$. Since $(r-2m -i\epsilon)^{-1}=P[(r-2m)^{-1}]+i\pi\delta(r-2m)$, $P$ denoting the principal part, we get 
\beq
\Im I=\frac{\pi E}{2\kappa}
\eeq
$\kappa=(4m)^{-1}$ being the horizon surface gravity. Identifying the emission probability with $P_{{\rm em}}\sim\exp(-2\Im I)=\exp(-\pi \kappa^{-1}E)$ gives then the temperature $\kappa/\pi$, twice the Hawking result. This is in common  with the Kerr metric, which  in Boyer-Lindquist coordinates reads
\beq
ds^2&=&-\frac{\Delta\rho^2}{\sigma}dt^2+\frac{\rho^2}{\Delta}dr^2+\rho^2d\theta^2 \\
&+&\frac{\Sigma\sin^2\theta}{\rho^2}(d\phi-\omega dt)^2\nonumber
\eeq                                        
with the standard definitions
\beq
\Delta &=& r^2-2mr+a^2, \quad \rho^2=r^2+a^2\cos^2\theta,  \\ 
\Sigma &=& (r^2+a^2)^2-\Delta a^2\sin^2\theta, \quad  \omega=\frac{2mar}{\Sigma}
\eeq
The parameter $a$ is the angular momentum per unit mass, $J=ma$, and $m$ is the ADM mass. The horizon is the largest root of $\Delta=0$: $r_{\pm}=m\pm\sqrt{m^2-a^2}$. The HJ equation is separable in the Kerr field \cite{Carter:1968ks}, i.e. $I=-Et+J\phi +W(r)+S(\theta)$. Solving the equation gives
\beq
W(r)=\int\!\!\!\!\!\!\!\searrow\sqrt{{\cal R}}\,\Delta^{-1}dr
\eeq
where ${\cal R}=[E(r^2+a^2)-aJ]^2-K\Delta$ and $K$ is a constant. The integrand has a simple pole at $r=r_+$, where $\Delta=0$. Therefore, as before,
\[
\Im I= \frac{\pi {\cal R}(r_+)}{r_+-r_-}=\frac{\pi(r_+^2+a^2)}{r_+-r_-}(E-\Omega J)=\frac{\pi(E-\Omega J)}{2\kappa}
\]
where $\kappa$ is the surface gravity and $\Omega=a/2mr_+$ is the angular velocity of the horizon.  Again $P_{{\rm em}}\sim\exp(-2\Im I)=\exp[-\pi \kappa^{-1}(E-\Omega J)]$ gives $T=\kappa/\pi$, twice the Hawking result. 

A number of solutions were advanced. In \cite{Angheben:2005rm,Stotyn:2008qu} it was proposed to use the proper distance from the horizon, on the ground of covariant requirements. In \cite{Akhmedova:2008dz} it was convincingly proved that a similar imaginary part would be produced by the temporal part of the action (see also \cite{deGill:2010nb}). Others ventured to suggest that there is not even an imaginary part, the temporal part cancelling the simple pole on using the HJ equation \cite{Belinski:2009bc}. Of course in general there is also an amplitude to cross the horizon inward \cite{Mitra:2006qa}
\beq
\Im\int\!\!\!\!\!\!\!\!\nearrow dI, \qquad I=-Et-\int^r \frac{E rdr}{r-2m} + J\phi
\eeq
because to an outside observer the particle never reaches the horizon in  real Schwarzschild time. Hence deforming the contour  as above would give
\[
\Im\int\!\!\!\!\!\!\!\!\nearrow dI=-\frac{\pi E}{2\kappa}
\]
Therefore
\beq\label{main}
2\Im\int\!\!\!\!\!\!\!\searrow dI - 2\Im\int\!\!\!\!\!\!\!\!\nearrow dI=\frac{2\pi E}{\kappa}
\eeq
Taking the exponential gives
\beq\label{db}
P_{{\rm em}}=P_{{\rm abs}}e^{-2\pi E/\kappa}
\eeq
which is recognized as the detailed balance condition for a system in thermal bath. This is certainly correct but the derivation looks very suspicious. On the one hand the pole prescription seems contrived just to obtain the wanted result. On the other hand the use of singular coordinates to describe horizon crossing trajectories is very awkward, a point which was made clear by the Parikh and Wilczek treatment of the problem. We cover this last point with an example. The metric (\ref{sch}) can be made regular across the horizon by passing to Eddington-Finkelstein advanced coordinates $(v,r,\theta,\phi)$
\beq
ds^2=-\left(1-\frac{2m}{r}\right)dv^2 + 2dv dr +r^2d\omega^2
\eeq
The action is $I=-Ev+J\phi +W(r) + S(\theta)$ (note that $\partial_vI=\partial_tI$) and from the HJ equation, $g^{\mu\nu}\partial_{\mu}I\partial_{\nu}I=0$, one gets
\beq\label{ef}
W(r) = 2E\int^r \frac{r dr}{r-2m}
\eeq
Notice the factor two coming from the $2\partial_vI\partial_rI$ term in the HJ equation. On the ingoing path on the other hand one can easily see that there is no pole on crossing the horizon and consequently no imaginary part. From (\ref{ef}) we obtain the correct result
\beq
\Im\int\!\!\!\!\!\!\!\searrow dI=\frac{\pi E}{\kappa}
\eeq
Similar conclusions can be drawn using other regular coordinates, for  example the Painlev\'e-Gullstrand coordinates employed by Parikh \& Wilczek. The idea that by using coordinates which are regular across the horizon eliminates all sort of problems is one point made in \cite{DiCriscienzo:2010zza}.

We will soon see that Eq.~(\ref{main}) is correct (for Schwarzschild BH) although  not always the ingoing term adds an imaginary part, except that when it does the coordinates as a rule fail to cover the horizon.  Note that the left hand side of (\ref{main}) is a coordinate scalar, so it must be possible to obtain it from invariant arguments. This is provided by the analytic continuation of the classical action throughout complexified space-time.

\section{The analytic argument} 
\begin{figure}[h]
\includegraphics[scale=.4]{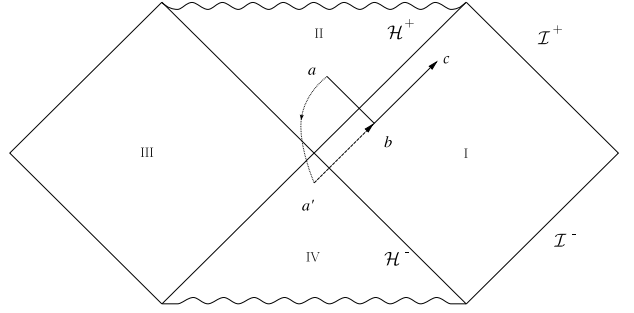}
\caption{Rotation of the tunneling path to cross the past horizon, where it becomes classically allowed. The path actually moves away from the plane of the figure.}
\label{fig}
\end{figure}

To justify the above machinery, and in particular the Feynman $i\epsilon$-prescription, we shall now rotates the tunneling path away from the horizon, as shown in Figure [\ref{fig}].

We shall use in intermediate steps the well known Kruskal coordinates $(U,V)$, such that $U<0$, $V>0$ in region $I$, $U>0$, $V>0$ in region $II$, $U<0$, $V<0$ in region $IV$,  and go to complex $(U,V)$-plane putting $\tilde{U}=U\exp(i\lambda)$, $\tilde{V}=V\exp(-i\lambda)$, $0\leq\lambda\leq\pi$. To understand this choice note that it corresponds to a Wick-like rotation of Schwarzschild time $t\to t-i\lambda/\kappa$. Now
\beq\label{cont}
dI&=&\partial_{\tilde{U}}Id\tilde{U}+ \partial_{\tilde{V}}Id\tilde{V}=\partial_{U}IdU+ \partial_VIdV \\
&+& [U\partial_UI-V\partial_VI]id\lambda
\eeq
where a term $\partial_{\phi}Id\phi +\partial_{\theta}Id\theta$ has been omitted since it does not give contributions to the imaginary part (but see the Kerr solution below). From the property of Kruskal coordinates one has
\beq\label{cont1}
-U\partial_UI+V\partial_VI=\kappa^{-1}\,\partial_tI=-\kappa^{-1}E
\eeq
Assuming analyticity, the integral over the segment $a\to b\to c$ is now equal to the integral over the semi-circle $[0,\pi]$ (over which $U$, $V$ are constant) plus the integral over the segment $a^{'}\to b\to c$, over which $\lambda$ is constant (and equal to $\pi$); thus we obtain 
\beq\label{main1}
\Im\int\!\!\!\!\!\!\!\searrow dI=\Im\int\!\!\!\!\!\!\!\nearrow dI+\frac{\pi E}{\kappa}
\eeq
where now the upward arrow refers to the path crossing the past horizon, which is classically allowed. Therefore $E$ is the conserved energy of the particle. The Feynman prescription is now clear, because it is the only one which is consistent with the analytic method. By time reversal invariance the amplitude to cross the past horizon outward is the same as the amplitude to cross the future horizon inward, therefore Eq.~(\ref{main1}) is just the same as  Eq.~(\ref{main}). If instead we choose to continue analytically the other way, say by putting $\tilde{U}=U\exp(-i\lambda)$, $\tilde{V}=V\exp(i\lambda)$, $0\leq\lambda\leq\pi$, which correspond to a counter clockwise Wick rotation, $t\to t+i\lambda/\kappa$ and a Feynman prescription $r\to r+i\epsilon$, we would obtain 
\beq\label{db0}
P_{{\rm abs}}=P_{{\rm em}}e^{-2\pi E/\kappa}
\eeq
This can be interpreted as the detailed balance condition for a white hole to absorb a quantum particle via the past horizon, a process that would be classically forbidden by causality. \\
{\it Spinning particles - } All we come to say should applies equally well to spinning particles. The known Lagrangian formulation  of such systems do not modify the free action  term, which is where the pole at the horizon resides. Equivalently, the Hamilton-Jacobi equation for fermions is just the same as for spinless particles as it represents  the phase of the spinor amplitude \cite{Kerner:2007rr,Vanzo:2008dm,Yale:2008kx}. It is important that these expectations were recently extended to spin-$1$ bosons, and that the Hawking temperature will not receive higher order corrections in $\hbar$ beyond the semi-classical ones \cite{Yale:2010tn} (see also \cite{Banerjee:2009wb} for a different view). 
\\
{\it Kerr black hole - } We can extend the calculation to the Kerr solution by noticing that throughout the complex manifold the azimuthal angle must also be rotated to keep the metric regular, more precisely $\phi\to \phi-i\Omega\lambda/\kappa$. Then adding the term $\partial_{\phi}Id\phi$ to the differential $dI$ would produce an imaginary term after a $\pi$-rotation, equal to $-i\pi\Omega\partial_{\phi}I/\kappa$. The outgoing trajectory from the past horizon is a classical solution so $\partial_{\phi}I=J$, the conserved angular momentum. We obtain in this way the result
\beq\label{main2}
\Im\int\!\!\!\!\!\!\!\searrow dI=\Im\int\!\!\!\!\!\!\!\nearrow dI+\frac{\pi}{\kappa} (E-\Omega J)
\eeq
or
\beq\label{db1}
P_{{\rm em}}=P_{{\rm abs}}e^{-2\pi( E-\Omega J)/\kappa}
\eeq
Of course in quantum theory the angular momentum is quantized. As is well known the emission and absorption probabilities for particles with energy $E$ and angular momentum $j$ are related to the Bogoliubov $\beta$-coefficients, whose computation is a classical problem involving the relevant field equations. Unitarity in the space of classical solutions relates them to the transmission coefficient $\Gamma_{Ejm}$ through the potential barrier surrounding the horizon
\beq\label{unit}
P_{{\rm abs}}\pm P_{{\rm em}}=\Gamma_{Ejm}
\eeq 
where the $(+)$ is for fermions and the $(-)$ for bosons. Together with (\ref{db}) it gives the Bose-Einstein or Fermi-Dirac spectrum.\\
{\it Charged black holes - } The prototypical charged solution is the Reissner-Nordstr\"o{}m metric
\beq
ds^2=-V(r)dt^2+\frac{1}{V(r)}dr^2+r^2d\omega^2
\eeq
where
\beq
V(r)=1-\frac{2m}{r}+\frac{q^2}{r^2}=\frac{1}{r^2}(r-r_+)(r-r_-)
\eeq
and the electromagnetic field has potential $A=r^{-1}Qdt$ (a one-form). The action to be integrated on the tunneling path is $dI_0 +eA$, where $I_0$ is the free action and $e$ the electric charge. The form $A$ is ill-defined at the horizon, for this reason one usually makes a gauge transformation to a form $\tilde A=A+df$ which is regular there. In our case the analytic continuation takes $A$ away from the horizon so this will actually be unnecessary. Using as above complex $(\tilde U,\tilde V)$ coordinates we obtain
\beq\label{main3}
\Im\int\!\!\!\!\!\!\!\searrow dI=\Im\int\!\!\!\!\!\!\!\nearrow dI+\frac{\pi}{\kappa} (E-e\Phi)
\eeq
where $E=-\partial_tI_0$ is the mechanical energy and $\Phi=q/r_+$. The quantity $E-e\Phi$ is gauge invariant and conserved along the outgoing path from the past horizon. The extension to cover the Kerr-Newmann solution should now be obvious.\\
{\it A generalized Parikh-Wilczek method - } The previous considerations suggest a simple generalization of the null geodesics method. The authors manage to compute the imaginary part of the integral of the radial momentum in Painlev\'e-Gullstrand coordinates, namely
\beq
\Im\int\!\!\!\!\!\!\!\searrow p_rdr
\eeq
This looks non covariant, but we may substitute the full Liouville differential one-form, $\varpi=p_{\mu}dx^{\mu}$, in place of the radial momentum, which is nothing but the reduced action. We can now analytically continue as explained above, first  by writing $\varpi=p_{U}dU+p_{V}dV$, then rotating $(U,V)$ from zero to $\pi$ and finally integrating along the rotated curve. As in Eq.~(\ref{cont}) the imaginary part will be $i\pi(Up_U-Vp_V)$: but, as in Eq.~(\ref{cont1}),  this is $-i\pi p_t/\kappa=i\pi E/\kappa$, where $E$ is the Killing energy as measured at infinity. In all we get
\beq\label{main4}
\Im\int\!\!\!\!\!\!\!\searrow\varpi-\Im\int\!\!\!\!\!\!\!\nearrow\varpi=\frac{\pi E}{\kappa}
\eeq
Incidentally this shows that the null geodesic method and the HJ method are completely equivalent as far as stationary black holes are concerned. A confirmation of this fact based on specific coordinate systems was presented in \cite{Kerner:2006vu}. This is true because the Hawking effect is an energy conserving process, so that the reduced action is all one needs in a static geometry. 

\section{Conclusions} 
The main result of this work, namely equations (\ref{main1}), (\ref{main2}), (\ref{main3}) and (\ref{main4}), show that it is possible to formulate a coordinate invariant statement about semi-classical horizon tunneling.  It is not even necessary to use the Hamilton-Jacobi equation or the Hamiltonian equations of motion, although one needs to know which paths are forbidden and which ones are not. Nor it is necessary to prescribe some special coordinate system, as sometime it is rumored in relation to the Painlev\'e-Gullstrand frame.  In particular the  imaginary temporal contributions can be present or not, depending on the chosen time, but they will never cancel the pole part.  In fact the formalism appears covariant and therefore independent on which particle concept (or time) one employs. It is hoped that this will contribute to a better understanding of the tunneling mechanism.\par
We wish to acknowledge R.~Di Criscienzo and G.~Acquaviva for useful discussions and for supporting the figures.

%% The Appendices part is started with the command \appendix;
%% appendix sections are then done as normal sections
%% \appendix

%% \section{}
%% \label{}

%% References
%%
%% Following citation commands can be used in the body text:
%% Usage of \cite is as follows:
%%   \cite{key}          ==>>  [#]
%%   \cite[chap. 2]{key} ==>>  [#, chap. 2]
%%   \citet{key}         ==>>  Author [#]

%% References with bibTeX database:

%\bibliographystyle{model1-num-names}
%\bibliography{<your-bib-database>}

%% Authors are advised to submit their bibtex database files. They are
%% requested to list a bibtex style file in the manuscript if they do
%% not want to use model1-num-names.bst.

%% References without bibTeX database:

\end{document}